\documentclass[a4paper,11pt]{article}
\pdfoutput=1 % if your are submitting a pdflatex (i.e. if you have
\usepackage{jheppub} % for details on the use of the package, please
  \usepackage{graphicx}                   % see the JHEP-author-manual
\usepackage{epstopdf}
\usepackage[T1]{fontenc} % if needed
\usepackage{amsmath}
\usepackage{slashed}
 \usepackage{graphicx}

\title{\boldmath Time-reversal symmetry violation in several 
Lepton-Flavor-Violating processes % An effective approach 
}

\author[a,b,c]{Juan Carlos Vasquez}
\affiliation[a]{ICTP, \\  Strada Costiera, 11  - 34151 Trieste Italy}
\affiliation[b]{SISSA/INFN, \\ via Bonomea, 265 - 34136 Trieste, Italy}
\affiliation[c]{Gran Sasso Science Institute, \\ Viale Crispi 7, 67100 
L'Aquila, 
Italy}

\emailAdd{jcvasque@sissa.it}

\abstract{We compute a  T-odd triple vector correlation for the  
$\mu\rightarrow e\gamma $ decay and the $\mu\rightarrow e$ conversion process,    
 finding simple results in terms of the CP violating phases of the effective 
Hamiltonians. Then  we  focus on the  minimal Left-Right symmetric extension of 
the 
Standard Model, which can lead to an appreciable correlation. We show that  under 
rather general assumptions, this correlation can be used to discriminate  between 
Parity or Charge-conjugation as the discrete Left-Right
symmetry.}

\begin{document} 
\maketitle
\flushbottom

\section{Introduction}
\label{sec:intro}
 Lepton Number Violating (LNV) and Lepton Flavor  violating (LFV) processes are 
forbidden  in the Standard Model (SM)  and are thus a good probe of new 
physics. 
In principle new physics brings also new sources of CP violation and therefore time reversal (T) symmetry violation in any local, Lorentz invariant  quantum field theory.

Motivated by this  we explicitly  compute   T-odd triple vector correlations for the LFV  
$\mu \rightarrow e\gamma$ decay  and $\mu\rightarrow e$ 
conversion process, since much of the present and future experimental 
efforts are devoted to these two processes. The MEG collaboration reports the best
experimental  limit for the $\mu\rightarrow e\gamma$ decay \cite{Adam:2013mnn}
\begin{equation}
\textbf{Br}(\mu\rightarrow e\gamma)\equiv\frac{\Gamma(\mu\rightarrow 
e\gamma)}{\Gamma(\mu\rightarrow e\nu_{\mu}\nu_e)} < 5.7\times 10^{-13}\label{MEGbound}
\end{equation}
and  the SINDRUM II collaboration gives the strongest limits for the 
$\mu\rightarrow e$ conversion process \cite{LFVlimit1,LFVlimit2}, namely
\begin{equation}
\textbf{Br}(\mu + \text{Ti} (\text{Au}))\rightarrow e+\text{Ti} (\text{Au}) )\equiv \frac{\Gamma(\mu\rightarrow e)}{\Gamma_{\text{capt}}}
<6.1(7)\times 10^{-13}, \label{SINDRUMII}
\end{equation}
 where $\Gamma_{\text{capt}}$ is the muon capture rate in the vicinity of a nucleus.
  Upgrades of ongoing experiments have been considered  with 
the final goal of achieving a sensitivity around $10^{-18}-10^{-19}$ 
\cite{deGouvea:2010zz,Miller,Witte,Mihara:2013zna}. Given the current  limits and the future 
improvements, there exist the possibility of having enough statistics to start 
probing CP violation beyond the SM in the next round of experiments. This is 
suggested and studied in \cite{Farzan:2007us,Ayazi:2008gk}.

In this work we focus on quantities that test T violation 
  in the absence of final-state interactions and among 
these quantities  are  
triple vector correlations made up of the momenta or spins of the participating 
particles \cite{Wolfenstein:1990ks}. In \cite{Bajc:2009ft}, it is suggested that  
triplet vector correlations can be used to probe CP violation in the 
$\mu\rightarrow e$ conversion process. Here we present the first 
analytical computation for the  correlation suggested in \cite{Bajc:2009ft} for 
the $\mu\rightarrow e$ conversion process and we extend their work in two ways: 
first,  we compute  the correlation for the $\mu\rightarrow e\gamma$ decay and  
second  we include the full set of effective operators that enter the 
$\mu\rightarrow e$ conversion process. 

In section \ref{sec2} we introduce some  theoretical tools for the 
$\mu\rightarrow e\gamma$ and 
the $\mu\rightarrow e$  conversion process. In section \ref{sec3} as an 
example of a theory 
 that gives  order one contribution to the triple vector
correlation, we briefly introduce the minimal Left-Right (LR) symmetric 
extension of the SM. In section \ref{sec4} and \ref{sec5}  we present the 
analytical 
computation of the triple correlation in the $\mu \rightarrow e\gamma$ and $\mu 
\rightarrow e$ conversion process respectively. Then in section \ref{sec6} and 
for both processes, we 
study these correlations in the context of the minimal LR 
model, for both parity and charge-conjugation as the LR symmetries. 
Finally in section \ref{sec7} we present our  conclusions.

% % % % % % % % % % % % % % % % %
\section{General theory \label{sec2}}
\subsection{ $\mu\rightarrow e\gamma$ process}

The $\mu  \rightarrow e \gamma $ decay  is predicted 
to be negligible small in the SM  with massive neutrinos, therefore if this 
process is seen it implies that  new physics is behind 
it.    The effective Hamiltonian for this process is of the form 
 \begin{equation}
 H_{eff}=\frac{4eG_Fm_{\mu}}{\sqrt{2}}\bar{e}(p_e)\sigma_{\mu 
\nu}F^{\mu \nu}(A_LP_L+A_RP_R)\mu(p_{\mu}) +h.c., \label{leff}
 \end{equation}
 where $e$ is the electromagnetic coupling constant,  $F_{\mu \nu}$ is the electromagnetic field strength for the photon 
field, $G_F$ is the Fermi constant, $P_{(R,L)}\equiv\frac{1}{2}(1\pm\gamma_5)$ 
, 
$m_{\mu}$ is the muon mass and 
 $e(p_e)$ and $\mu(p_{\mu})$ are the spinors for the electron and muon  respectively.
 For this process we use the gamma matrices in the Weyl basis and  the 
coefficients $A_L$ and $A_R$ are calculated within a given physical model.

\subsection{$\mu\rightarrow e$ conversion. Theory and Effective Hamiltonian }

 Theoretical studies of this process were performed  in the past 
\cite{Weinberg:1959zz,Marciano:1977cj,Shanker:1979ap,Kitano:2002mt}. In 
\cite{Kitano:2002mt} the outgoing electron coming from 
the conversion process,  belongs to one of the states in the continuum energy 
spectrum for the Coulomb potential and as a matter of fact the 
outgoing electron must be treated as a plane wave. One way to argue this  is 
by 
noticing that an 
electron in the continuum energy spectrum,   is described by a Dirac spinor in 
the angular momentum basis. Experimentally, the detected electron has a define 
4-momentum implying that the outgoing electron must be a plane wave.

In this work we  present a method for computing a triple vector  correlation that 
tests T-violation in the 
$\mu\rightarrow e$ conversion process for various nuclei. We make use of the 
formalism developed 
in \cite{Rose1952}.

We  use the following representation for the 
$\gamma$ matrices
\begin{eqnarray}
\gamma_0= \beta=
\left( \begin{array}{ccc}
1&&0 \\ 
0&&-1\\
\end{array} \right) , \qquad
\gamma_i= 
\left( \begin{array}{ccc}
0&&\sigma_i \\ 
-\sigma_i&&0\\
\end{array} \right),
\end{eqnarray}
and
\begin{align}
&\sigma_{\mu\nu}=\frac{i}{2}[\gamma_{\mu},\gamma_{\nu}], \qquad 
\gamma_5=-i\gamma_1\gamma_2\gamma_3\gamma_0,
\end{align}
where the 
$\sigma_i$ are the Pauli matrices where $i=1,2,3$ and the index $\mu$ takes the values $\mu =0,1,2,3.$

The Dirac's equation for the central field problem in polar coordinates  is 
given by (the energy is 
given in units of the electron mass)
\begin{equation}
E\psi=H \psi= [-i \gamma_5\Sigma_r(\frac{\partial}{\partial 
r}+\frac{1}{r}-\frac{\beta}{r}K)+V+\beta]\psi, \label{3}
\end{equation}
where
\begin{align}
&\Sigma_r=\frac{1}{r}\sum_i\Sigma_i, \quad
\Sigma_i= \frac{i}{2}[\gamma_j,\gamma_k] \qquad \text{(\{i,j,k\} cyclic)}. \\
& K=\beta(\Sigma \cdot L +1).
\end{align}
 $V$ is the Coulomb potential and $L$ is the orbital angular momentum.

We  write the wave function  as \cite{RoseBook} 
\begin{eqnarray}
 \psi_{\kappa}^{\mu}=
\left( \begin{array}{ccc}
g_{\kappa}(r)\chi_{\kappa}^{\mu} \\ 
if_{\kappa}(r)\chi_{-\kappa}^{\mu}\\
\end{array} \right),\label{spinorcentral}
\end{eqnarray}
such that $K \psi_{\kappa}^{\mu}=-\kappa \psi_{\kappa}^{\mu}$ and $J_3 \psi_{\kappa}^{\mu}=\mu  \psi_{\kappa}^{\mu}$, where $J_3$ is the third component of the total angular momentum $\vec{J} $.
The radial functions $g_{─\kappa}$ and $f_{\kappa}$ obey the 
differential equations
 
\begin{align}
&\frac{dg_{\kappa}(r)}{dr}=-\frac{\kappa+1}{r}g_{\kappa}(r)+(E-V+1)f_{\kappa}
(r)\label{diffeqn1}
, \\ 
&\frac{df_{\kappa}(r)}{dr}=\frac{\kappa-1}{r}f_{\kappa}(r)-(E-V-1)g_{\kappa}
(r).\label{diffeqn2}
\end{align}

 In the high energy limit -all the masses are set to zero- and  from 
eqs.\eqref{diffeqn1} and \eqref{diffeqn2},  $f_{\kappa}(r)$ and $g_{\kappa}(r)$ 
satisfy
 \begin{equation}
 f_{-\kappa} = -g_{\kappa}, \qquad g_{-\kappa}=f_{\kappa}.
 \end{equation}

From here on we make  use of  this result for the spinor $\psi_{\kappa,E}^{\mu(e)}$ describing the  electrons  coming from 
the conversion process. The initial muon instead is described by $\psi_{\kappa}^{\mu}$  with the quantum numbers, $\mu=\pm\frac{1}{2}$ 
and $\kappa=-1$  and we choose the normalization 
\begin{equation}
\int d^3x\psi_{1s}^{(\mu)\dagger}(\vec{x})\psi_{1s}^{(\mu)}(\vec{x})=1.
\end{equation}

For the  electrons in  the continuum-energy states we  use the same 
normalization considered
in \cite{Kitano:2002mt}, namely
\begin{equation}
\int d^3x 
\psi_{\kappa,E}^{\mu(e)\dagger}(\vec{x})\psi_{\kappa^{'},E^{'}}^{\mu^{'}(e)}
(\vec{x})=2\pi\delta_{\mu 
\mu^{'}}\delta_{\kappa^{'}\kappa}\delta(E-E^{'}).\label{normalizationelectron}
\end{equation}

In the conversion process the effective Hamiltonian is 
given by \cite{Kitano:2002mt}
\begin{align}
& H_{eff}=\frac{4G_F}{\sqrt{2}}(m_{\mu}A^*_R\bar{\mu}\sigma^{\mu 
\nu}P_LeF_{\mu \nu}+m_{\mu}A^*_L\bar{\mu}\sigma^{\mu \nu}P_ReF_{\mu \nu}+h.c.) 
\nonumber \\
&+\frac{G_F}{\sqrt{2}}\sum_{q=u,d,s}[(g_{LS(q)}\bar{e}P_R\mu+g_{RS(q)}\bar{e}
P_L\mu)\bar{q}q+(g_{LP(q)}\bar{e}P_R\mu+g_{RP(q)}\bar{e}P_L\mu)\bar{q}\gamma_5q 
\nonumber \\
&(g_{LV(q)}\bar{e}\gamma^{\mu}P_L\mu+g_{RV(q)}\bar{e}\gamma^{\mu}P_R\mu)\bar{q}
\gamma_{\mu}q+(g_{LA(q)}\bar{e}\gamma^{\mu}P_L\mu+g_{RA(q)}\bar{e}\gamma^{\mu}
P_R\mu)\bar{q}\gamma_{\mu}\gamma_5q+ \nonumber \\
&\frac{1}{2}(g_{LT(q)}\bar{e}\sigma^{\mu \nu}P_R\mu+g_{RT(q)}\bar{e}\sigma^{\mu 
\nu}P_L\mu)\bar{q}\sigma_{\mu \nu}q + h.c.].\label{Heffm2e}
\end{align}
 The nuclear form factors were calculated in \cite{Kosmas:2001mv}. The wave 
function for the muon and the electrons in the presence of a central field were 
obtained in \cite{Shanker:1979ap,Kitano:2002mt}. In particular in 
\cite{Kitano:2002mt} updated data for the proton and neutron 
densities were used.

In the limit of  $r\rightarrow \infty$ it can be shown that the general solution  for a Dirac 
particle in a Coulomb field at first order in $H_{eff}$ is of the form 
\cite{Rose1952}
\begin{eqnarray}
	\psi_{as}= -i\sqrt{\frac{\pi}{|\vec{p}|}}\frac{e^{ipr}}{r}\sum_{\kappa 		
\mu}e^{i\delta_{\kappa}}\langle\psi_{\kappa}^{(e)\mu}|H_{eff}|\psi^{(\mu)}_{1s} 
	\rangle
	\left( \begin{array}{ccc}
		\sqrt{E+1}\chi_{\kappa}^{\mu}(\hat{p}) \\ 
		-\sqrt{E-1}\chi_{-\kappa}^{\mu}(\hat{p})\\
	\end{array} \right)+\mathcal{O}(H^2_{eff}), \label{electronout}
\end{eqnarray}
where $\hat{p}$ is in the direction of the outgoing electron.
 The phases $e^{i\delta_{\kappa}}$ are the usual ones  appearing in scattering 
problems in the presence of a Coulomb field and are given by
\begin{align}
	&\delta_{\kappa}=y\ln2pr-\arg\Gamma(\gamma+iy)+\eta_{\kappa}-\frac{1}{2}
	\pi\gamma  \label{deltak},\\
	& y=\alpha Z E/p, \quad \gamma=\sqrt{\kappa^2-\alpha^2Z^2}, \quad 
	e^{2i\eta_{\kappa}}=-\frac{\kappa-iy/E}{\gamma+iy}
\end{align}
where  $Z$ is the atomic number, $\alpha=e^2/4\pi$ and $p$ is the modulus of the 3-momentum $\vec{p}$. We consider states with $\kappa=\pm1$, hence the only term relevant for our 
discussion is
$\eta_{\kappa}$ --the  remaining ones are just an overall phase 
in the solution $\psi_{as}$.

 Finally the total conversion rate per unit flux is  \footnote{See appendix 
\ref{appendixB} for a more detailed discussion on this issue.}
\begin{equation}
\omega_{conv} = R^2\int d\Omega  \psi_{as}^{\dagger} 
\psi_{as}=\frac{1}{2}\sum_{\kappa,\mu}|\langle\psi_{\kappa}^{\mu}|H_{eff}|\psi_i \rangle|^2.
\end{equation}

\section{The minimal Left-Right symmetric theory\label{sec3}}

As an example of a complete and predictive theory of lepton number violating phenomena   we consider the minimal LR symmetric extension of the SM~\cite{Pati:1974yy,lrmodel,Mohapatra:1974gc,spont}. In this model the  gauge group is  $SU(2)_L\times SU(2)_R\times U(1)_{B-L}$ with an additional discrete symmetry that may be generalized parity ($P$) or  charge conjugation ($C$) --for reviews  see \cite{Senjanovic:2010nq,Senjanovic:2011zz,tesisvladimir}. It relates  the smallness of neutrino masses  to the near maximality of parity violation al low energies through the seesaw mechanism \cite{Minkowski:1977sc,Mohapatra:1979ia, Mohapatra:1980yp,seesawothers,seesawothers2}.
The  scalar 
sector contains the following  fields \cite{Minkowski:1977sc,GoranNuclphys79,Mohapatra:1979ia,Mohapatra:1980yp} 
 \begin{eqnarray}
 \Phi =
 \left( \begin{array}{ccc}
 \phi_1^0 && \phi_2^+  \\
 \phi_1^- && \phi_2^0 \\
 \end{array} \right) , \quad  \Delta_{L,R}=\left( \begin{array}{ccc}
 \delta^+_{L,R}/\sqrt{2} && \delta_{L,R}^{++}  \\
 \delta_{L,R}^{0} &&- \delta^+_{L,R}/\sqrt{2}  \\
 \end{array} \right), \nonumber \\
 \end{eqnarray}
 where $\Phi$ is in the (2,2,0) representation of  $SU(2)_L \times 
SU(2)_R \times U(1)_{B-L}$ and the  two scalar triplets $\Delta_L$ and 
$\Delta_R$, belong to the (3,1,2) and the (1,3,2) representations respectively.  The 
Yukawa interactions of leptons with the scalar triplets  have the form
\begin{eqnarray}
\mathcal{L}_Y=&\bar{L}_L(Y_{\Phi}\Phi+\tilde{Y}_{\Phi}\tilde{\Phi})L_R 
+\frac{1}{2}(L_L^TCi\sigma_2Y_{\Delta_L}\Delta_LL_L \nonumber \\ 
&+L_R^TCi\sigma_2Y_{\Delta_R}\Delta_RL_R )+h.c., \label{yukawas}
\end{eqnarray}
 $\tilde{\Phi}=\sigma_2\Phi^*\sigma_2$,  
$L_L$ is the lepton doublet of the standard model ($L_L^T=(\nu\quad l)_L$) and
$L_R$ is its right-handed analogue that we denote as  $L_R^T=(N\quad l)_R$ where $N$ is the heavy Majorana 
neutrino. The $Y_a$ is the Yukawa coupling of the field $a$, where $a=\{\Phi,\tilde{\Phi},\Delta_L,\Delta_R\}$.

 Under the discrete left-right symmetry the fields of the theory transform as:
 \begin{align}
 P:  \left\{ \begin{array}{ll}
 \mathcal{P}f_L\mathcal{P}^{-1} =  \gamma_0f_R  \\
 \mathcal{P}\Phi\mathcal{P}^{-1} =\Phi^{\dagger} \\
 \mathcal{P}\Delta_{L,R}\mathcal{P}^{-1} = -\Delta_{R,L}  
 \end{array} \right., \quad C :  \left\{ \begin{array}{ll}
 \mathcal{C}f_L  \mathcal{C}^{-1} =\boldsymbol{C}(\bar{f_R})^T  \\
 \mathcal{C}\Phi\mathcal{C}^{-1} = \Phi^{T} \\
 \mathcal{C}\Delta_{L,R}\mathcal{C}^{-1} = -\Delta^*_{R,L} 
 \end{array} \right.  \label{relations}
 \end{align}
  where  the usual charge conjugation operator is given by $\boldsymbol{C}=i \gamma_2\gamma_0$.
  
   Invariance of the Lagrangian under the LR symmetry leads to the following relations between the Yukawa couplings of the theory, namely 
    \begin{align}
     P :  \left\{ \begin{array}{ll}
     Y_{\Delta_{R,L} } = Y_{\Delta_{L,R}} \\
     Y_{\Phi}=Y_{\Phi}^{\dagger} \\
     \tilde{Y}_{\Phi}=\tilde{Y}_{\Phi}^{\dagger}
     \end{array} \right.,\quad
     C :  \left\{ \begin{array}{ll}
     Y_{\Delta_{R,L} } = Y^*_{\Delta_{L,R}} \\
     Y_{\Phi}=Y_{\Phi}^T \\
     \tilde{Y}_{\Phi}=\tilde{Y}_{\Phi}^T
     \end{array}. \right.
     \end{align}

In the mass eigenstate basis the flavor changing charged current Lagrangian is 
given by
\begin{equation}
	\mathcal{L}_{cc}= \frac{g}{\sqrt{2}}(\bar{\nu}_L V_{L}^\dag 
\slashed{W}_{L} l_L+\bar{N}_R V_{R}^\dag \slashed{W}_{R} l_R) 
+h.c.,\label{cclagrangian}‏
\end{equation}
 $V_R$ is the right-handed analogue of the PMNS mixing matrix $V_L$. In general 
it  has 
 three different mixing angles and six arbitrary complex phases and we
 parametrize it as
 \begin{equation*}
 	V_R= K_e \hat{V}_R K_N,
 \end{equation*}
 with $K_e\equiv\text{diag}(e^{i\phi_e},e^{i\phi_{\mu}},e^{i\phi_{\tau}})$, 
 $K_N\equiv\text{diag}(1,e^{i\phi_{2}},e^{i\phi_{3}})$. The matrix  $\hat{V}_R$ has  
three 
 mixing angles and the dirac phase $\delta$.  We choose for $\hat{V}_R$ the 
 standard form for the CKM matrix shown in the PDG \cite{Agashe:2014kda}.

The interaction terms of charged leptons with the doubly-charged scalars are
\begin{align}
&\mathcal{L}_{\Delta} = \frac{1}{2} l_R^TC Y'_{\Delta_R} \delta_R^{++}l_R +
\frac{1}{2} l_L^TC Y'_{\Delta_L} \delta_L^{++}l_L +h.c., \label{Ldelta} \\
& Y'_{\Delta_R} = \frac{g}{M_{W_R}}V_R^*M_NV_R^{\dagger}. \label{Y_R}
\end{align}

If charge conjugation  is the discrete LR symmetry,  the charged 
lepton masses are symmetric and in this case the Yukawa couplings in \eqref{Ldelta} satisfy  
(for  reviews on this topic see references  
\cite{Senjanovic:2010nq,Senjanovic:2011zz,tesisvladimir})
\begin{equation}
 Y'_{\Delta_L} = (Y'_{\Delta_R})^* . \label{Crelation}
\end{equation}

 For parity   and in the more interesting phenomenological 
situations, the charged lepton masses matrices are almost hermitian 
\cite{vladimirgoranmiha}. In \cite{Vasquez:2014mxa} it was realized that it 
implies the near equality between the Yukawa couplings  shown in Eq. \eqref{Ldelta} i.e. 
\begin{equation}
Y^{'}_{\Delta_L}=Y^{'}_{\Delta_R} +\mathcal{O}(\tan 2\beta \sin 
\alpha).\label{relationPcase}
\end{equation} 
The vacuum expectation values of the neutral fields belonging to $\Phi$ are such that $\langle \phi_1^0\rangle=v_1 $ and $\langle \phi_2^0\rangle=v_2 e^{i\alpha}$, where   $\beta$ is the ratio $v_2/v_1$ and $\alpha$ is the 
spontaneous phase.  In \cite{Maiezza:2010ic,Senjanovic:2014pva} it is shown 
that 
$\tan 2\beta\sin\alpha \lesssim 2m_b/m_t$  ($m_b$  and $m_t$  are the bottom and top quark  masses respectively), so that the Yukawa coupling of the 
doubly charged scalar are nearly equal \cite{Vasquez:2014mxa}. 
 
 It is a remarkable feature of the minimal LR theory, that  the TeV energy 
scale 
accessible at the LHC through the  Keung-Senjanovi\'c (KS) process \cite{Keung:1983uu} --and its 
associated LNV and LFV, predicts the  rate for the neutrino-less double beta 
decay and low energy LFV. This deep connection and the related phenomenology 
are 
illustrated in \cite{Tello:2010am,Nemevsek:2011aa}. All these processes depend in a crucial way 
on the elements of the leptonic right-handed mixing matrix $V_R$, for which all 
its mixing angles, the Dirac phase and two Majorana phases can be  determined 
at 
the LHC \cite{Vasquez:2014mxa}. Useful information can also be obtained from EDM of the neutron and such \cite{Zhang:2007da,An:2009zh,Xu:2009nt,Seng:2014pba,Bsaisou:2014zwa}. This is deeply connected to the study of the strong CP parameter, which in the mLRSM turns out to be calculable \cite{Mohapatra:1978fy,Maiezza:2014ala,Senjanovic:2015yea}. 

Recently the CMS collaboration \cite{Khachatryan:2014dka} has reported an excess in the ee-channel for this process at 2.8$\sigma$, but they claimed that this excess cannot be accommodated in the minimal version of the theory --assuming diagonal mixing in the right-handed leptonic sector and degenerate masses for the heavy neutrinos.  Several works  have been proposed \cite{Aguilar-Saavedra:2014ola,Aguilar-Saavedra:2014ola1,Heikinheimo:2014tba,Gluza:2015goa} in order to explain this excess and  the conclusion was that it  would need a higher Left-Right symmetry breaking scale, or  a more  general  mixing scenario with pseudo-Dirac heavy neutrinos.

 % % % % % % % % % % % % % % % % % % % % % % % % % % % % % % % %
\section{Computation of a triple vector correlation in the $\mu \rightarrow 
e\gamma$ decay\label{sec4}}

\begin{figure}
	\begin{center}
		\includegraphics[width=0.8\textwidth]{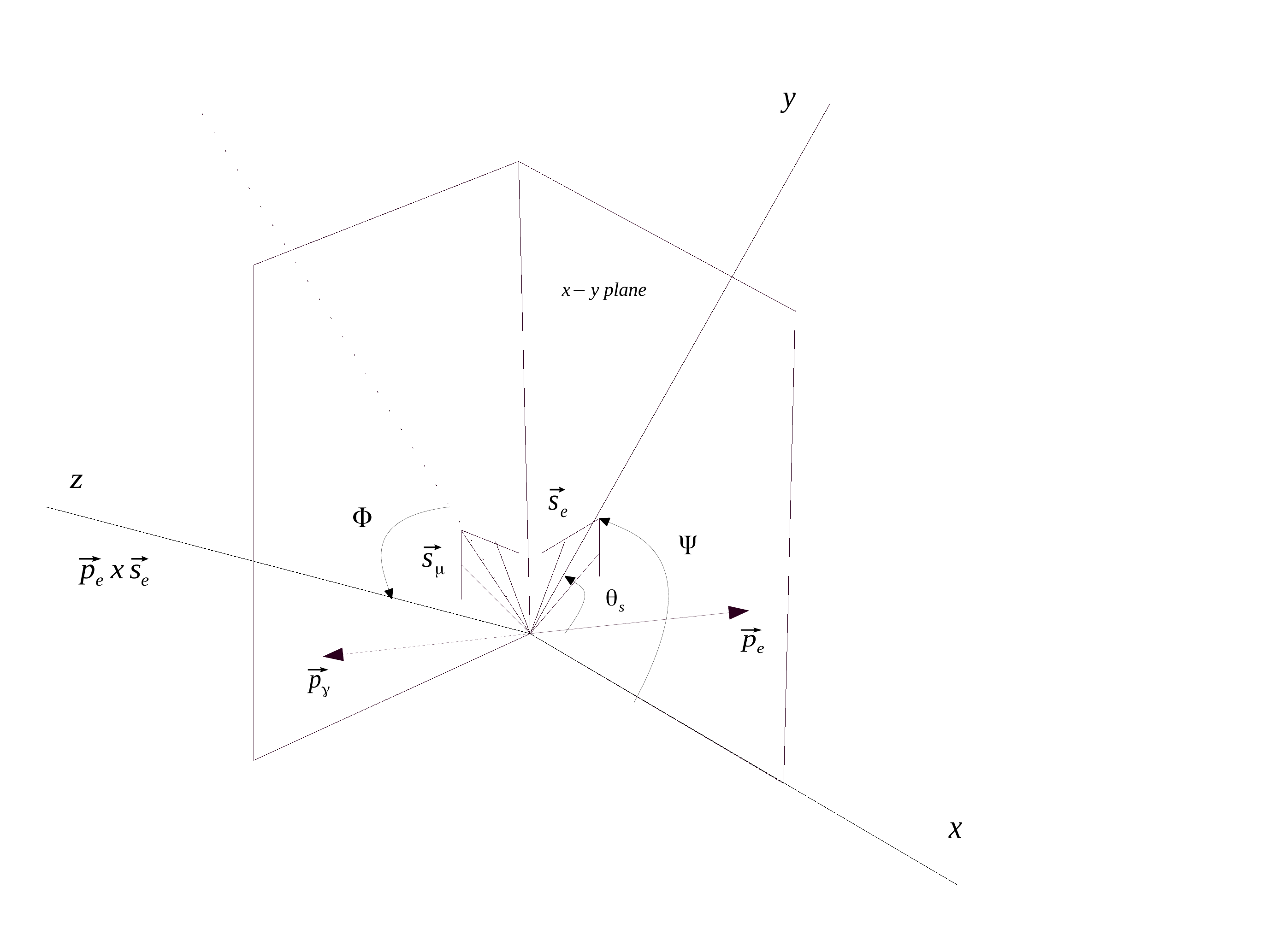} 
	\end{center}
	\caption{ \label{fig1}Reference frame and the setup for the 
		$\mu \rightarrow e \gamma$ decay.} 
\end{figure}

$T$-odd asymmetries in the $\mu \rightarrow e \gamma$ were considered in the 
past. 
In \cite{Farzan:2007us,Ayazi:2008gk}, it was shown that by 
studying the polarization of electron and the photon coming from the muon decay 
it is 
possible to extract the CP-violating phases  from the experiment. The 
conclusion 
was that in order to extract the CP-violating phases 
both electron  and photon polarizations must be measured. In this paper   instead, we present an alternative way of extracting the 
CP-violating phases of the effective Hamiltonian in the $\mu\rightarrow 
e\gamma$ 
decay.  This is complementary to the  work presented in 
\cite{Farzan:2007us,Ayazi:2008gk}. The novelty is that no measurements of the 
final photon polarizations are needed. We consider 
the T-violating triple vector product 
\begin{equation}
\hat{s}_{\mu^{+}} \cdot(\hat{p}_{e^{+}}\times\hat{s}_{e^{+}}) =  \cos \Phi \sin 
\theta_s,
\end{equation} 
where $\theta_s$ is the angle between the polarization's direction 
($\hat{s}_{e^+}$) of the positron  and its momentum's direction
$\hat{p}_{e^+}$, $\Phi$ is the angle formed between $\hat{s}_{\mu^+}$ and the 
direction 
defined by $\vec{p}_{e^+}\times \vec{s}_{e^+}$ and $\Psi$ is the azimuthal 
angle. In  Fig.\ref{fig1}  the reference frame and 
 setup are shown.  Notice that this quantity changes sign under parity and 
naive time-reversal transformation  $\hat{T}$  defined by  $t \rightarrow 
-t$. For processes whose interactions are characterized by a small coupling, 
it can be  shown    at first order that the connected part 
of the S-matrix is hermitian \cite{Wolfenstein:1990ks} and   therefore
the violation of the $\hat{T}$ symmetry amounts the violation of the 
time-reversal symmetry.

 We  define the triple vector correlation as
\begin{align}
&\langle  \hat{s}_{\mu^{+}} \cdot(\hat{p}_{e^+}\times\hat{s}_{e^{+}}) 
\rangle_{\Phi} 
\equiv \frac{N(\cos \Phi >0)-N(\cos \Phi<0)}{N_{total}}= \label{asymmetrygamma} 
\\  
\nonumber
&\frac{\int_0^\pi d\Phi  d \Gamma/d \Phi \cdot 
\text{sgn}(\hat{s}_{\mu^{+}}\cdot(\hat{p}_e\times 
\hat{s}_{e^{+}}))}{\Gamma_{total}} ,
\end{align}
where $\Gamma_{\text{total}}$ and  $N_{total}$ are  the total decay  rate and the total number of events for the initially polarized muon respectively, $N(\cos\Phi> 0)$ and $N(\cos\Phi< 0)$ are the number of events satisfying $\cos\Phi 
> 0$ and $\cos\Phi 
< 0$ respectively.

The 4-momenta of the participating particles in the rest frame of the muon  are given by
\begin{align}
	&p^{\mu}_{\mu^+}=(m_{\mu},0,0,0)\label{momentummuon}, \\
	&p^{\mu}_{e^+}=(E_e,|\vec{p}_{e^+}|\sin \theta_s,|\vec{p}_{e^+}|\cos 
\theta_s,0) \label{momentumpositron}, \\ 
	&p^{\mu}_{\gamma}=(E_{\gamma},-|\vec{p}_{e^+}|\sin 
\theta_s,-|\vec{p}_{e^+}|\cos \theta_s,0) \label{momentumphoton}
\end{align}
where the mass of the positron has been neglected. The energy $E_{e^+}$ of the 
positron and the 
energy $E_{\gamma}$ of the photon are given by
\begin{equation}
E_{e^+}\cong E_{\gamma}=|\vec{p}_{e^+}|=\frac{m_{\mu}}{2}.
\end{equation}

From the effective Hamiltonian in eqn. \eqref{leff} and  eqns. 
\eqref{spinormuon}, \eqref{spinelectron0} and \eqref{photon} in appendix 
\ref{appendixA},
  a straightforward computation gives the following value for the correlation
\begin{equation}
\langle  \hat{s}_{\mu^{+}} \cdot(\hat{p}_{e^+}\times\hat{s}_{e^{+}}) 
\rangle_{\Phi}=\sin \theta_s\frac{ \Im m(A_LA_R^*)}{|A_L|^2+|A_R|^2}.
\label{asym2eg} 
\end{equation}
  The main advantage of this 
quantity is that no measurements of the photon polarizations  are needed.

In summary we find that given a source of  polarized anti-muons, by measuring 
the 3-momentum $\vec{p}_{e^+}$ of the outgoing positron and its polarization 
$\vec{s}_{e^+}$,  the asymmetry shown in eqn. \eqref{asym2eg} is 
sensitive to the CP-violating phases of the effective Hamiltonian shown in \eqref{leff}. In 
\cite{Corriveau:1983cf,Burkard:1985wn,Kohler:2001ca,Fetscher:2003ke,
Fetscher:2003vu} it is shown  that measurements of the  polarization of 
electrons coming from the muon  decay are feasible. We assume a 100 $\%$ 
polarized muon flux so  that  our results must be trivially rescaled by the actual 
polarization of the initial muons.

% % % % % % % % % % % % % % % % % % % % % % % % % % % % % % % % % % % % % % %
% % % % % % % % % % % % % % % % % % % % % % % % % % % % % % % % % % % % % % %
\section{Computation of a triple vector correlation in the $\mu \rightarrow e$ 
conversion process\label{sec5}}
Following the same lines of  the last section, we   
define an asymmetry given by comparing the number of events with $ 
\vec{s}_{\mu}\cdot(\vec{p}_e\times \vec{s}_e)>0$ with the ones satisfying 
$ \vec{s}_{\mu}\cdot(\vec{p}_e\times \vec{s}_e)<0$  in the  $\mu \rightarrow e$ 
conversion process and  we define it as
\begin{align}
\langle  \hat{s}_{\mu} \cdot(\hat{p}_e\times\hat{s}_{e}) \rangle_{\Phi} 
& \equiv \frac{N(\cos \Phi >0)-N(\cos \Phi<0)}{N_{total}} \label{asymmetrymec} 
\nonumber 
\\
&= \frac{\omega_{conv}(\cos \Phi>0)-\omega_{conv}(\cos 
\Phi<0)}{\omega_{conv}}
\end{align}
where $\omega_{conv}$ is the total conversion rate and as previously, $\Phi$ is the angle between the plane formed by the vectors 
$\hat{p}_e $ and $\hat{s}_e$ and the polarization of the muon $\hat{s}_{\mu}$.
 We used the same  coordinate system shown in Fig.\ref{fig1} but clearly  there 
is no photon coming from the muon decay.

A direct computation of the asymmetry shown in Eq. \eqref{asymmetrymec} gives \footnote{For more  details see section \ref{projectionoperators}.}
\begin{align}
	\langle  \hat{s}_{\mu} \cdot(\hat{p}_e\times\hat{s}_{e}) \rangle_{\Phi}  
	=
	\frac{1}{2}\sin \theta_s\frac{ \Im m 
		(C_LC_R^*)}{|C_L|^2+|C_R|^2}+\mathcal{O}(\alpha 
Z)+\mathcal{O}(\frac{m_e}{E_e}). 
	\label{asymu2e}
\end{align}
 $m_e$ is the electron mass and 
\begin{align}
&C_R\equiv 
DA_R+S^{(p)}(\tilde{g}_{LS}^{(p)}+\tilde{g}_{LV}^{(p)})+S^{(n)}(\tilde{g}_{LS}^{
(n)}+\tilde{g}_{LV}^{(n)}), \\
&C_L\equiv 
DA_L+S^{(p)}(\tilde{g}_{RS}^{(p)}+\tilde{g}_{RV}^{(p)})+S^{(n)}(\tilde{g}_{RS}^{
(n)}+\tilde{g}_{RV}^{(n)})
\end{align}
where
\begin{align}
& \tilde{g}^{(p)}_{LS,RS} \equiv \sum_q G^{(q,p)}g_{LS,RS(q)},
& \tilde{g}^{(n)}_{LS,RS} \equiv \sum_q G^{(q,n)}g_{LS,RS(q)}, \\
&  \tilde{g}^{(p)}_{LV,RV} \equiv 2g_{LV,RV(u)} + g_{LV,RV(d)}, &  
\tilde{g}^{(n)}_{LV,RV} \equiv g_{LV,RV(u)} + 2g_{LV,RV(d)}.
\end{align}

$D$, $S^{(n,p)}$ are nuclear constants already calculated and tabulated in 
\cite{Kitano:2002mt} for various elements. $G^{(q,p)}$ and $ G^{(q,n)}$ are obtained from the scalar  matrix element  \cite{Kosmas:2001mv,Kitano:2002mt}
\begin{equation}
\langle N | \bar{q}q |N \rangle= Z G^{(q,p)}\rho^{(p)}+(A-Z) G^{(q,n)}\rho^{(n)}
\end{equation}
 $Z$ and $A$ are the atomic and mass number respectively, $\rho^{(n)}$ and $\rho^{(p)}$ are the neutron and proton densities inside the nucleus. 
 The  expression obtained is valid  for  
non-relativistic muons and 
 we droped 
terms of the order $\alpha Z$ and  $m_e/E_e$. In practice 
equation \eqref{asymu2e} must be multiplied by the  polarization of the initial 
muons, which  is of the order of $15 \%$  in the conversion process 
\cite{Mann:1961zz}.

\section{Triplet vector correlation in the minimal Left-Right 
theory\label{sec6}}

As a concrete example of a theory beyond the SM that gives order one values
for the T-odd triple vector correlation \cite{Bajc:2009ft} we consider the minimal LR 
symmetric 
model. In what follows we  analyze separately the contributions to the 
asymmetries \eqref{asym2eg} and \eqref{asymu2e} in the case of $P$ and $C$ as 
the LR symmetries.  In \cite{Bajc:2009ft} it is found that this contribution 
can 
be of order one, since  there are new contributions coming from interactions of 
charged 
leptons with the singly-charged and doubly-charged scalar fields.

\subsection{$\mu\rightarrow e\gamma$ decay}
In this section and for the $\mu\rightarrow e \gamma $ decay, we study the contributions to  the triple vector correlation for both Parity and Charge Conjugation as the LR symmetry. 

\textbf{ Parity as the LR symmetry:} in \cite{Cirigliano:2004mv} the authors presented a complete study 
of the 
contributions to several LFV 
processes in the context of the minimal LR extension of the SM and it is found that the  branching ratio for this process is of the form 
\begin{equation}
\text{Br}(\mu\rightarrow e \gamma) = 384 \pi^2 e^2 (|A_L|^2+|A_R|^2)
\end{equation}
where
\begin{align}	
&A_R=\frac{1}{16\pi^2}\sum_n(V^{\dagger}_R)_{en}(V_R)_{n\mu}[\frac{M_{W}^2}{M^2_		
{W_R}}S_3(X_n)-\frac{X_n}{3}\frac{M^2_{W}}{M^2_{\delta^{++}_R}}], \label{A_R}\\	
&A_L=\frac{1}{16\pi^2}\sum_n(V^{\dagger}_R)_{en}(V_R)_{n\mu}X_n[-\frac{1}{3}	
\frac{M^2_{W}}{M^2_{\delta^{++}_L}}-\frac{1}{24}\frac{M^2_{W}}{M^2_{H_1^{+}}}] 
	+\mathcal{O}(\tan 2\beta \sin \alpha) \label{A_L},\\ 
	&X_n = (\frac{M_N}{M_{W_R}})^2, \quad 	
S_3(x)=-\frac{x}{8}\frac{1+2x}{(1-x)^2}+\frac{3x^2}{4(1-x)^2}[\frac{x}{(1-x)^2}
	(1-x+\log x)+1].
\end{align}
  $M_{N_n}$ are the heavy neutrino masses where  $n=1,2,3$. $M_W$ is the W boson mass, $M_{W_R}$ is the $W_R$ boson mass, $M_{H_1^+}$ is the  mass of the heavy scalar $H_1^+$ and  $M_{\delta_{(L,R)}^{++}}$ are the masses for the left and right doubly charged scalars respectively and finally we use $M_{\nu}$ to denote the light neutrino masses.

\begin{figure}
	\centering
	\includegraphics[width=0.5\textwidth]{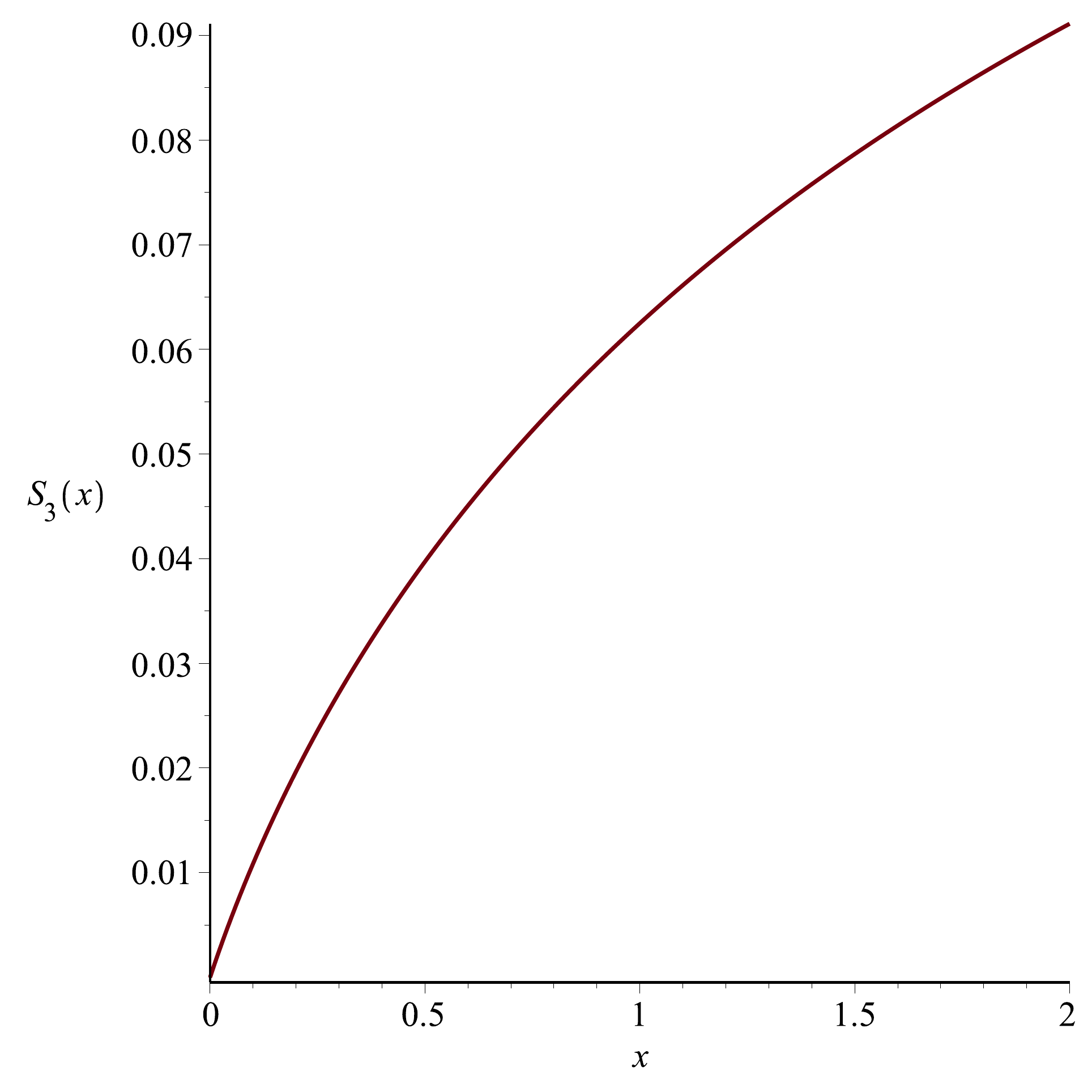} 
	\caption{\label{fig3} Plot of the loop function $S_3(x)$.}
\end{figure}

Notice that the loop function $S_3$ is always small as far as  $M_N$ is not much 
bigger than $M_{W_R}$, so that the term with the loop function  can 
neglected for a wide range of the heavy neutrino masses (see figure \ref{fig3}) and therefore 
  the correlation defined in \eqref{asym2eg} is suppressed.  
 Finally we neglect 
the contribution of the charged Higgs $H_1^{+}$ since its mass  cannot be lower 
than (15-20) TeV \cite{Maiezza:2010ic,Bertolini:2014sua}. This poses no problem 
for the theory, since its mass emerges at the large scale of symmetry breaking 
\cite{GoranNuclphys79,Senjanovic:1979cta}.
The gauge boson and doubly-charged scalar masses can be obtained at the LHC 
through the so called KS process and the decays of the doubly charged scalars 
\cite{Keung:1983uu} in addition with  all the  mixing angles and the 
Dirac phase in $V_R$
 \cite{Vasquez:2014mxa}.  This is an  example 
of  the complementary role played by the  high and low energy 
experiments in the establishment of the LR theory 
\cite{Tello:2010am,Chakrabortty:2012pp,Chakrabortty:2012mh,Deppisch:2012nb,
Barry:2013xxa,Dev:2013oxa,Dev:2014xea}.  

For the sake of illustration, imagine that type II see-saw is the dominant 
source of neutrino masses i.e. $\frac{M_N}{\langle \Delta_R \rangle} = 
\frac{M_{\nu}}{\langle \Delta_L 
	\rangle}$ and $V_L=V_R$. In this case it 
is possible to show that the heavy neutrino masses satisfy the relation  \cite{Tello:2010am}
\begin{equation}
\frac{M^2_{N_2}-M^2_{N_1}}{M^2_{N_3}-M^2_{N_1}}= 
\frac{M^2_{\nu_2}-M^2_{\nu_1}}{M^2_{\nu_3}-M^2_{\nu_1}} \simeq \pm 0.03,
\end{equation}
\begin{figure}
	\centering
	\includegraphics[width=6.2in]{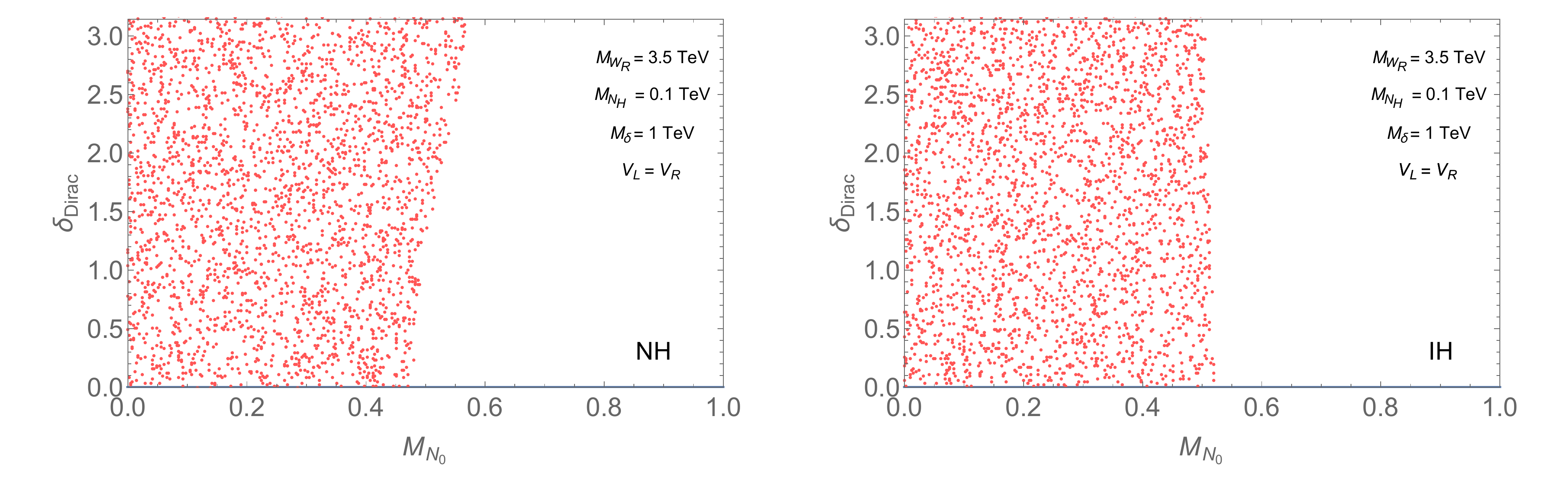}\\
	\includegraphics[width=6.2in]{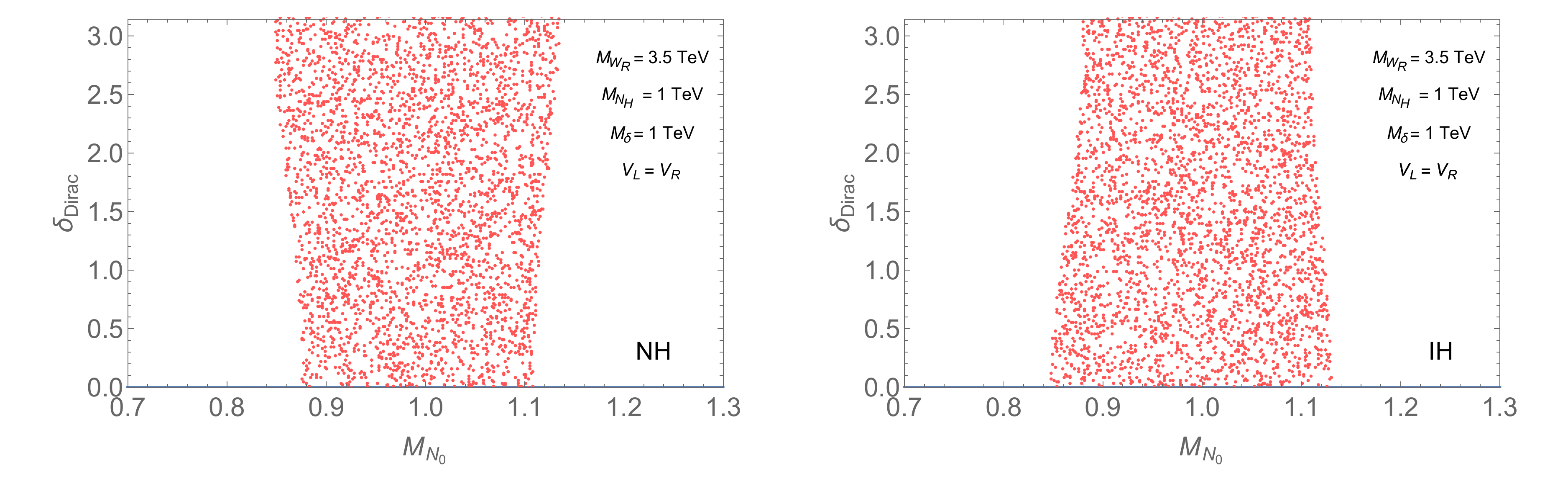} 
	\caption{\label{fig5}  Plot obtained by considering the MEG bound shown in Eq. \eqref{MEGbound}. (Right) Normal hierarchy case (NH). (Left) Inverse hierarchy case (IH). The colored region is the allowed one. (Top) Mass of the heaviest right-handed neutrino $M_{N_H} = 0.1 $ TeV. (Bottom)  Mass of the heaviest right-handed neutrino $M_{N_H} = 1 $ TeV. }
\end{figure}
where the $\pm$ corresponds to normal (NH) and inverted (IH) neutrino mass hierarchy  respectively. In what follows we denote  $M_{N_0}$  the lightest right-handed
neutrino mass,  $M_{N_H}$ the heaviest right-handed neutrino mass
and $\delta$ is the Dirac phase present in $\hat{V}_R$.  In Fig. \ref{fig5} and for the two representative values of  $M_{N_H}= 0.1$ TeV and $M_{N_H}= 1$ TeV  we show  the allowed region obtained from the MEG bound in the $\{M_{N_0},\delta_{\text{Dirac}}\}$ plane,  for both normal and inverted  neutrino mass spectrum. The region between these values gives rise to the exciting LNV signals at the LHC trough  the KS process. We assume $M_{W_R}=3.5$ TeV  and common masses for the doubly charged scalars $M_{\delta_L^{++}}=M_{\delta_R^{++}}=M_{\delta}=1$ TeV.  The reader may ask about the very different behavior obtained for the two values of the heaviest neutrino mass chosen, and the point is that this can be  readily understood by noticing that the amplitude is approximately proportional to $|\Delta M_{13}^2|=|M_{N_H}^2-M_{N_0}^2|$, so that a bound  is obtained for $|\Delta M_{13}^2|$ rather on the lightest neutrino mass itself.

\begin{figure}	
	\centering
	\includegraphics[width=7in]{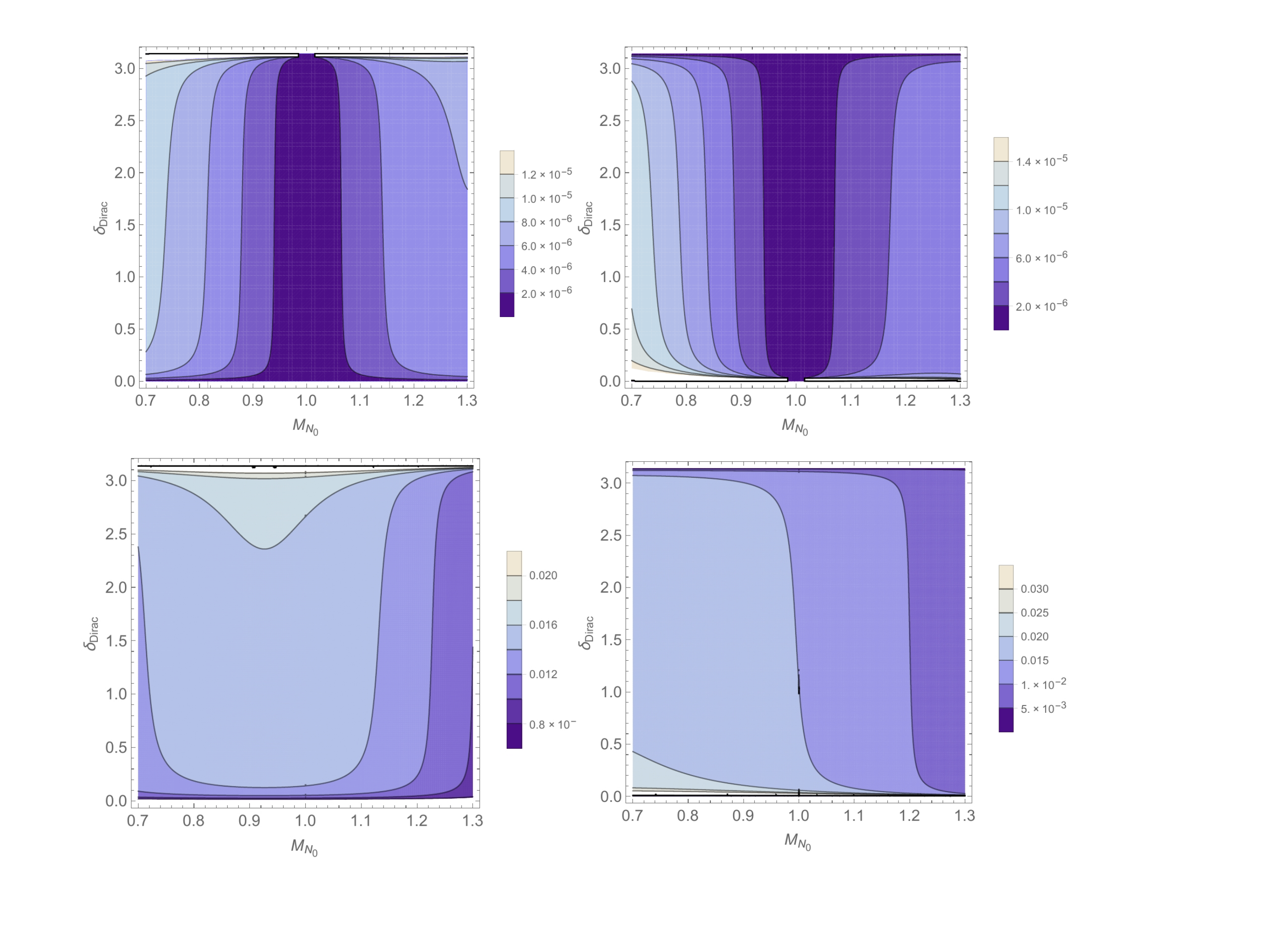}
	\caption{\label{fig4} (Top) Contour plots illustrating the absolute value of the 
		asymmetry defined in \eqref{asym2eg} as a function of the 
		lightest neutrino mass 
		$M_{N_0}$ and the Dirac phase $\delta$ for $P$ as the LR symmetry. 
		(Bottom) Contour plots illustrating the value of the asymmetry 
		defined in 
		\eqref{asym2eg} as a function of the lightest neutrino mass 
		$M_{N_0}$ and the Dirac 
		phase $\delta$ (assuming $\phi_{\mu}-\phi_e=0$) for $C$ as the 
		LR symmetry.
		(Left) Normal hierarchy for neutrino masses. (Right) Inverse 
		hierarchy 
		for neutrino masses.  We take 
		the gauge boson mass  $M_{W_R}=3.5$TeV, the heaviest right-handed neutrino 
		mass 
		$M_{N_H}=1$TeV and common  masses 
		for the doubly charged scalars of $M_{\delta}=1$ TeV. The mixing angles are 
		$\theta_{12}\simeq33.6^o$, $\theta_{23}\simeq41.9^o$, $\theta_{13}\simeq8.7^o$.}
\end{figure}

In 
figure \ref{fig4} (top) we plot the absolute  value for  the  triple vector correlation  given in 
\eqref{asym2eg} in the $(M_{N_0},\delta)$-plane, where one may  see that  the values of the correlation 
\eqref{asym2eg} goes from $10^{-6}$ to $10^{-5}$ in the allowed region.

One would be tempted to conclude that  the triple vector correlation may be bigger for general values of neutrino masses and 
mixings. However from eqns. 
\eqref{relationPcase},  the 
contribution to  the triple vector correlation shown in \eqref{asym2eg} is bounded to be less 
$10^{-2}$ since 
$\tan2\beta\sin\alpha<10^{-2}$ from the quark masses 
\cite{Maiezza:2010ic,Senjanovic:2015yea,Senjanovic:2014pva}. The point is that for  charged leptons masses ($M_l$)  bigger or equal than the Dirac mass of neutrinos ($M_D$), the mass matrix of the charged leptons is nearly hermitian leading therefore to nearly equal leptonic left and right mixing matrices. This is  in complete analogy to the situation in the quark sector studied in \cite{Zhang:2007da,Maiezza:2010ic}. Of course it is  possible to assume that $M_D>M_l$, but we will not pursue this possibility since in this case the original see-saw mechanism would lose its meaning and one would have to invoke  accidental cancellations in order to explain the smallness neutrino masses.

\textbf{Charge conjugation as the LR symmetry}: from eqn.
\eqref{Crelation} we have that
\begin{align}	
&A_R=\frac{1}{16\pi^2}\sum_n(V^{\dagger}_R)_{en}(V_R)_{n\mu}[\frac{M_{W}^2}{M^2_		
{W_R}}S_3(X_n)-\frac{X_n}{3}\frac{M^2_{W}}{M^2_{\delta^{++}_R}}], \label{A_R1}\\	
&A_L=\frac{1}{16\pi^2}\sum_n(V^{T}_R)_{en}(V^{*}_R)_{n\mu}X_n[-\frac{1}{3}\frac{		
M^2_{W}}{M^2_{\delta^{++}_L}}-\frac{1}{24}\frac{M^2_{W}}{M^2_{H_1^{+}}}]  
	\label{A_L1}.
\end{align}

Notice that some of  the external phases appearing in $V_R$ do not cancel in 
\eqref{asym2eg} and the triple vector correlation is proportional to 
$e^{2(\phi_{\mu}-\phi_{e})}$, so that  the triple vector correlation is not suppressed by 
the small $\theta_{13}$ mixing-angle. In Fig.\ref{fig4} (bottom) we show the absolute
value of the triple vector correlation in the $(M_{N_0},\delta)$-plane. We take 
$(\phi_{\mu}-\phi_{e})=0$ in
both normal and inverted  neutrino mass hierarchies. For $(\phi_{\mu}-\phi_{e})=\pi/4$ it will reach in 
maximum value of around $0.5$ in almost all the parameter space

Finally from Fig.\ref{fig4} (bottom)  we conclude that  $C$ as the LR 
symmetry   gives  larger contributions to the triple vector correlation and this  because in the parity case, the triple vector correlation  is 
suppressed due to the near equality between the Yukawa couplings.

The bottom line is that in the most interesting region of the parameter space,  a value for the triple vector correlation bigger 
than 
$10^{-2}$ can only be the consequence of $C$ as the LR 
symmetry.

  One may ask whether  this value for the asymmetry of could be measured in forthcoming experiments. Suppose that $\mu\rightarrow e\gamma$ is found to be of the order of $10^{-14}$. In the  best scenario due to the future experimental improvements on the sensitivity,  it would become possible  to observed at most $10^{4}$ events and  out of these  events one has to select the ones that have $\theta_s \neq 0 $ or $\theta_s \neq \pi$. Moreover  suppose that only the events satisfying  $\pi/6 < \theta_s< \pi/3$ may be identify in the experiment due to its intrinsic sensitivity. This would imply that we end up having $10^{4} \int_{\pi/6}^{\pi/3} \sin \theta_s d\theta_s \sim 10^3 $ events in the most optimistic situation. Hence this naive argument allow us to conclude that in most optimistic scenario,  an asymmetry of the order $10^{-3}$ or bigger would  probably be seen in the next round of $\mu\rightarrow e \gamma$ decay experiments.

\subsection{$\mu\rightarrow e$ conversion process}
In this section we consider the triple vector  correlation  for the 
$\mu\rightarrow e$ conversion process in the context of the minimal LR 
symmetric 
extension of the SM where the relevant branching ratio is given by \cite{Cirigliano:2004mv}
\begin{align}
\text{Br}(\mu \rightarrow e )= \frac{2G_F^2 V^{(p)2}}{\Gamma_{\text{capt}}}(\frac{\alpha^2}{16\pi^2})\left(|F^{(\gamma)}_L|^2+|F^{(\gamma)}_R|^2\right).
\end{align}
 The values of the capture rate  $\Gamma_{\text{capt}}$ are tabulated in \cite{Suzuki:1987jf} for several elements. In \cite{Cirigliano:2004mv} it was shown that the contribution of the 
 doubly-charged scalar may dominate due to a 
 logarithmic enhancement and in this case  the functions $F^{(\gamma)}_L$ and $F^{(\gamma)}_R$ may be written as 
 \begin{equation}
 	F^{(\gamma)}_{(L,R)} \simeq  128 \pi^2 A_{(L,R)}\log (m_{\mu}^2/ M^2_{\delta_{(L,R)}^{++}}). \label{logenhance}
 \end{equation}
 \begin{figure}
 	\centering
 	\includegraphics[width=6.2in]{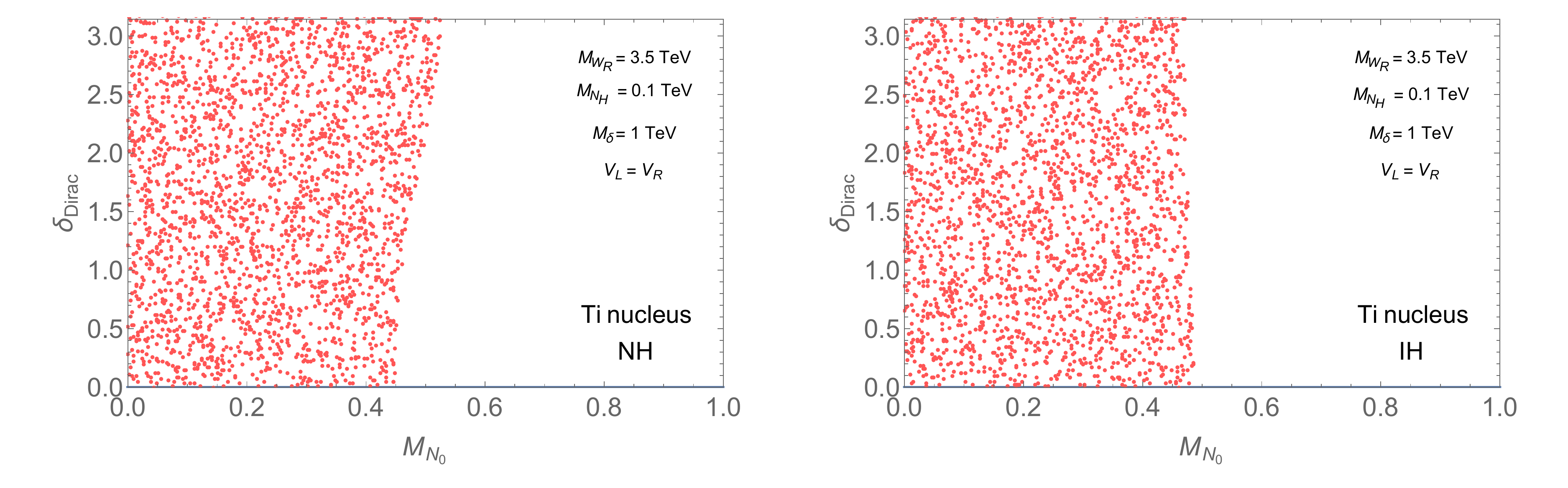} \\
 	\includegraphics[width=6.2in]{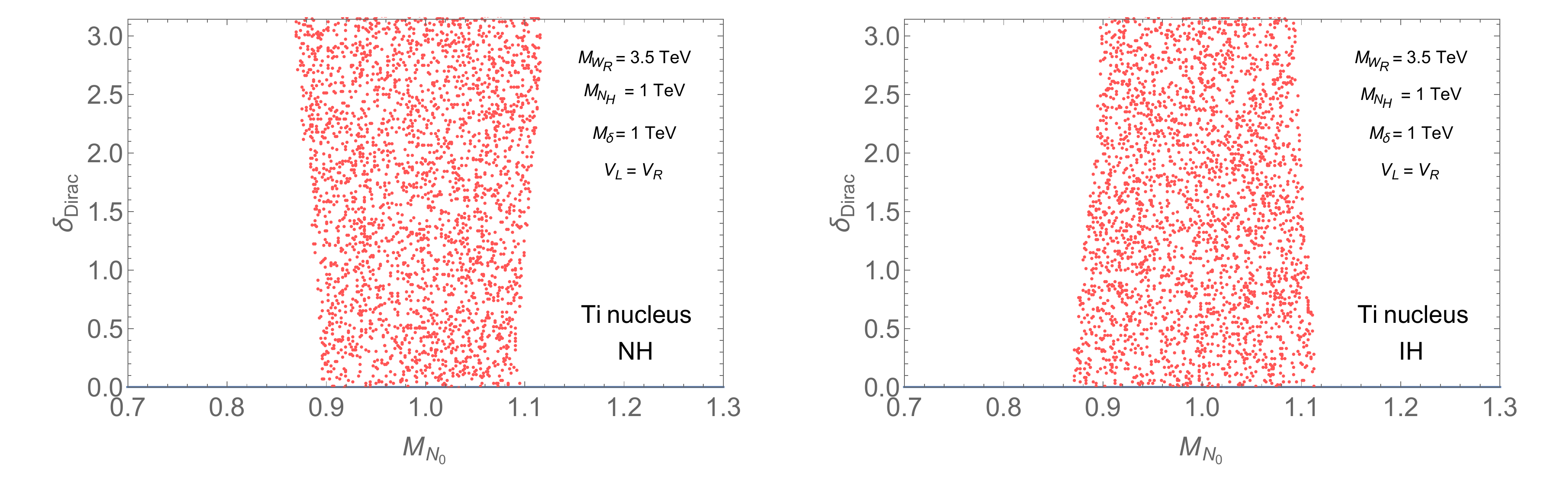} 
 	\caption{\label{fig6}  Plot obtained by considering the SINDRUM II bound for Titanium shown in Eq. \eqref{SINDRUMII}. (Right) Normal hierarchy case (NH). (Left) Inverse hierarchy case (IH). The colored region is the allowed one.  (Top) Mass of the heaviest right-handed neutrino $M_{N_H} = 0.1 $ TeV. (Bottom)  Mass of the heaviest right-handed neutrino $M_{N_H} = 1 $ TeV.  }
 \end{figure}
For completeness we show in Fig. \ref{fig6} the allowed region obtained by considering the SINDRUM bound for Titanium shown in Eq. \eqref{SINDRUMII} assuming the same   values for the heavy neutrino masses of the last section. From Eq. \eqref{logenhance} and assuming that the dominant terms are the logarithmic enhance ones, the amplitude for the conversion process and the $\mu\rightarrow e \gamma $ decay are proportional. Therefore a similar  qualitative behavior is obtained.      We can see that the bound obtained is similar to the one of the $\mu \rightarrow e \gamma$ experiment and this is due to the fact that the logarithmic enhancement in Eq. \eqref{logenhance} compensate the $\alpha$ suppression  in the conversion rate  \cite{Cirigliano:2004mv}.  For Gold  the bound one would obtained is  similar since  the ratio between the  conversion rates for the two elements is around 0.83.  On the other hand, for the gold atom relativistic effects of the muon becomes relevant, so that the result shown in Eq. \eqref{asymu2e} cannot be trusted in this case.

 Finally   the asymmetry defined in  Eq. \eqref{asymu2e} takes the form
\begin{equation}
\langle\vec{s}_{\mu}\cdot(\vec{p}_e\times \vec{s}_e)\rangle_{\Phi}= 
  \frac{\sin\theta_s}{2}\frac{\Im 
m(F^{(\gamma)}_LF^{*(\gamma 
		)}_R)}{|F^{(\gamma)}_L|^2+|F^{(\gamma)}_R|^2}=\frac{\sin \theta_s}{2}\frac{ \Im m(A_LA_R^*)}{|A_L|^2+|A_R|^2},
\end{equation}
where it can be   seen  that  this asymmetry has the same flavor 
structure 
of the coefficients $A_L$ and $A_R$ defined previously for the $\mu \rightarrow e \gamma$ decay, therefore the same 
conclusion obtained in the $\mu\rightarrow e\gamma$ case holds for the 
$\mu\rightarrow e$ conversion process as well. % The factor of 2 in the denominator is due to the extra sum over $\kappa$ when computing  the total conversion rate.

Regarding the expected sensitivity for the conversion process the arguments we used in the $\mu \rightarrow  e \gamma $ decay  apply, but with the difference that the final sensitivity is rescaled by a factor of the order of $10^{-1}$ due to the depolarization  --around $15\%$-- of the muons in the conversion process \cite{Mann:1961zz}. 

\section{Conclusions \label{sec7}}
We derived analytical expressions for a T-odd triple vector correlation in the 
$\mu\rightarrow e\gamma$ decay and the $\mu\rightarrow e$ conversion process 
and 
 found  simple results in terms  of the CP-violating phases of the 
effective 
Hamiltonians. The expression obtained in the $\mu\rightarrow e$ conversion omits 
relativistic corrections for the muons, but is otherwise complete. For the 
$\mu\rightarrow e \gamma $ decay we conclude that in order to extract the CP 
violating phases of the theory from the experiment, no measurements of the 
photon polarizations are needed. 

Then as an example of a theory that leads  order one  values for the triple vector correlation 
we consider the TeV scale, minimal Left-Right symmetric extension of the SM. 
Remarkably, due to the relation  between left and right Yukawa couplings in 
\eqref{Ldelta} --see also eqs. \eqref{Crelation} and \eqref{relationPcase}-- 
this triple vector correlation can be used to discriminate between charge-conjugation  or 
parity 
 as the Left-Right symmetry. More precisely, if the Dirac masses of  neutrinos  are smaller than the masses of the charge leptons,  a value for the  triple vector correlation 
bigger than $10^{-2}$ can only be the consequence of 
charge-conjugation as the Left-Right symmetry.

\appendix
\section{Kinematics of the $\mu\rightarrow e\gamma$ process and the triple vector correlation\label{appendixA}}
In this appendix we give some  tools that could be useful when computing the triple vector correlation shown in Eq. \eqref{asym2eg} for
the $\mu\rightarrow e\gamma$ decay.

 For the anti-muon we use the spinor  $v(p_{\mu^+})$  given by 
\begin{eqnarray}
	v(p_{\mu^+}) =
	\left( \begin{array}{ccc}
		\sqrt{p\cdot \sigma} \quad \xi \\ 
		-\sqrt{p \cdot \bar{\sigma}} \quad\xi\\
	\end{array} \right), \label{spinormuon}
\end{eqnarray}
where $\xi^{\dagger} \xi=1$ and $p_{\mu^+}$ is given in Eq. \eqref{momentummuon}. As shown in Fig. \ref{fig1} the polarization vector of the muon is given by:
\begin{equation}
	\vec{s}=|\vec{s}|(\sin \Phi \cos \Psi,\sin \Phi \sin \Psi,\cos \Phi) 
\label{spin}
\end{equation} 
and it is straightforward to show that in this case
\begin{align}
	\xi^{n}=   \left( \begin{array}{ccc}
		e^{-i\frac{\Psi}{2}} \cos \frac{\Phi}{2}  \\ 
		e^{i\frac{\Psi}{2}} \sin \frac{\Phi}{2} \\
	\end{array} \right).
\end{align}
 One may find the same result by requiring $\xi$ to be an 
eigenvector of $\vec{\sigma}\cdot \hat{n} $, where $\hat{n}$ is a unitary 
vector 
in the direction of $\vec{s}$.

 For the electron and for the reference frame shown in Fig.\ref{fig1} we use
 \begin{eqnarray}
 v_{e^+}(p_{e^+}) =
 \sqrt{\frac{|\vec{p}_{e^+}|}{2}} \left( \begin{array}{ccc}
 -2e^{i\frac{\theta_s}{2}}\sin \frac{\theta_s}{2}  \\ 
 2ie^{-i\frac{\theta_s}{2}}\sin \frac{\theta_s}{2}  \\
 2ie^{i\frac{\theta_s}{2}}\cos \frac{\theta_s}{2}\\
 -2e^{-i\frac{\theta_s}{2}}\cos \frac{\theta_s}{2}
 \end{array} \right). \label{spinelectron0}
 \end{eqnarray}

 The  photon has two  possible polarizations along the direction 
of motion and in the particular frame we are considering in Fig.\ref{fig1} its 
polarization 
vector is given by,
 \begin{eqnarray}
 \epsilon_{\pm}^{\mu}(p_{\gamma}) =
 \frac{1}{\sqrt{2}} \left( \begin{array}{ccc}
 0 \\ 
 \pm i\cos \theta_s\\
 \mp i \sin \theta_s\\
 1
 \end{array} \right)  \label{photon}
 \end{eqnarray}
 where we can explicitly see that when $\theta_s=0$, the photon can only have a 
polarization $\pm 1$ along the y-axis and $p_{\gamma}$ and $p_{e^+}$ are the 4-momentum of the outgoing photon and electron respectively -- see Eq. \eqref{momentumpositron} and \eqref{momentumphoton}. Once the expressions for the spinors of the participating fermions  and the polarization vector of the photon are known, it is easy straightforward to compute the triple vector asymmetry given in \eqref{asym2eg}. 

We found that the total decay rate  is given by 

\begin{equation}
	\Gamma_{\text{total}}= \frac{2}{\pi}G_F^2m_{\mu}^5e^2(|A_L|^2+|A_R|^2).
\end{equation}
 
It would be interesting to compare the above equation with   the result one gets when summing the decay rates for $\cos\Phi>0$ to that of $\cos\Phi<0$, namely
\begin{equation}
\Gamma(\cos\Phi>0)+\Gamma(\cos\Phi<0)= 	\frac{2}{\pi}G_F^2m_{\mu}^5e^2(\cos^2\frac{\theta_s}{2}|A_L|^2+\sin^2\frac{\theta_s}{2}|A_R|^2).
\end{equation}
 On the other hand, by subtracting the total decay rates  for $\cos\Phi>0$ to that of $\cos\Phi<0$ one gets:
 \begin{equation}
 \Gamma(\cos\Phi>0)-\Gamma(\cos\Phi<0)=\frac{2}{\pi}G_F^2m_{\mu}^5e^2 \sin\theta_s \Im m (A_LA_R^*)
 \end{equation}
from which the asymmetry shown in \eqref{asym2eg} can be readily computed.
% % % % % % % % % % % % % %
\section{$\mu\rightarrow e$ total conversion rate and the triple vector correlation\label{appendixB}}

\subsection{Total conversion rate }

In this appendix we briefly comment about the amplitude of the $\mu\rightarrow 
e$ conversion process and  the Born's approximation we used.

In computing the $\mu\rightarrow e$ conversion process, one usually  assumes 
the 
so called Born's approximation for the outgoing electrons. This approximation 
has 
two meanings: one  is computing 
the conversion rate to a given order in  some small coupling; and the other  is 
the assumption that electrons coming from the conversion process are plane 
waves. The point is that we can do  better  and have a complete control of both 
approximations at the same time. More precisely for the relativistic 
one-electron atom and in the limit of  big $r$ ($r >> 
r_0$, where $V(r\geq r_0)=0$), the solution of the Dirac's equation at first 
order in the perturbation $H_{eff}$ is  of the form \cite{Rose1952}
\begin{eqnarray}
\psi_{as}= -i\sqrt{\frac{\pi}{|\vec{p}|}}\frac{e^{ipr}}{r}\sum_{\kappa 
\mu}e^{i\delta_{\kappa}}\langle\psi_{\kappa}^{\mu}|H_{eff}|\psi_i \rangle
\left( \begin{array}{ccc}
\sqrt{E+1}\chi_{\kappa}^{\mu}(\hat{p}) \\ 
-\sqrt{E-1}\chi_{-\kappa}^{\mu}(\hat{p})\\
\end{array} \right)+\mathcal{O}(H^2_{eff}),
\end{eqnarray}
where $\psi_i$ is any stationary state of the Coulomb potential, 
$\psi_{\kappa}^{\mu}$ is one of the continuum energy solutions and $H_{eff}$ is 
the effective Hamiltonian for the $\mu\rightarrow e$ conversion process. 
Furthermore it 
can be shown that  $\psi_{as}$ is an eigenfunction of $\vec{\alpha}\cdot 
\vec{p} 
+\beta$ with eigenvalue $E$ so that $\psi_{as}$ describes, indeed a plane wave \cite{Rose1952}.
In the high energy limit --neglecting the electron mass-- the solution 
$\psi_{as}$ simplifies to
\begin{eqnarray}
\psi_{as}= -i\sqrt{\pi}\frac{e^{ipr}}{r}\sum_{\kappa 
\mu}e^{i\delta_{\kappa}}\langle\psi_{\kappa}^{\mu}|H_{eff}|\psi_i \rangle
\left( \begin{array}{ccc}
\chi_{\kappa}^{\mu}(\hat{p}) \\ 
-\chi_{-\kappa}^{\mu}(\hat{p})\\
\end{array} \right).
\end{eqnarray}
Finally if  we are interested in computing the total conversion amplitude per 
unit  flux  (for a detector placed at fixed radius $r=R$) the total conversion 
rate is given by
\begin{equation}
\omega_{conv} = R^2\int d\Omega  \psi_{as}^{\dagger} 
\psi_{as}=2\pi\left(\frac{1}{2}\sum_{\kappa,\mu}|\langle\psi_{\kappa}^{\mu}|H_{eff}|\psi_i \rangle|^2\right)= 2 G_F^2(|C_L|^2+|C_R|^2)
\end{equation}
where the coefficients $C_L$ and $C_R$ are defined in section \ref{sec5} and   we may absorb  the $\sqrt{2\pi}$ factor into  the normalization of the wave 
function $\psi_{\kappa}^{\mu}$ in order to agree with the conventions adopted in \cite{Kitano:2002mt}.

\subsection{Triple vector correlation in the conversion process \label{projectionoperators}}

 In this appendix we give details of the calculation for the triplet correlation asymmetry in the $\mu \rightarrow  e $ conversion process within the formalism developed in \cite{Rose1952}. 
Since we are interested in describing  particles with  a given  polarization,   we are going to make use of the spin projection operators for Dirac spinors.  Instead of using the  covariant spin projection operator we make use of  the following   projection operator 

\begin{equation}
 P_{\hat{n}_0}^{(\pm)}= \frac{1}{2}(1\pm\mathcal{O}\cdot \hat{n}_0),
\end{equation}
where
\begin{equation}
 \mathcal{O}\equiv\beta \vec{\sigma}+(1-\beta)(\vec{\sigma}\cdot \hat {p})\hat{p}
\end{equation} 
  and $\hat{n_0}$ is the direction of the spin polarization vector in the rest frame of  the particle, $\hat{p}$ is the direction of its momentum and the $\pm$ represent positive and negative polarization respectively. Furthermore it can be shown that the description of the spin with this operator is equivalent to the usual one given by the manifestly covariant spin operator \footnote{ see \cite{RoseBook} chapter III.}. Notice that the non-relativistic limit of can be taken in a transparent way by replacing $\beta \rightarrow 1$. 

For our present problem we assumed the muon to be  non-relativistic  and in the frame shown in Fig.\ref{fig1} its polarization vector is of the form

\begin{equation}
 n_{\mu}= (0,\hat{n}_0),
\end{equation}
where 
\begin{equation}
 \hat{n}_0=(\sin \Phi\cos \Psi, \sin \Phi \sin \Psi, \cos \Phi)
\end{equation}
 by multiplying the  wave function of the muon in the conversion process by $P_{\hat{n}_0}^{(+)}$ one obtains the wave function of a non-relativistic muon with the given  polarization. For the electron instead a full relativistic treatment is required since its energy is  $E_e= m_{\mu}-\epsilon_b$, where $m_{\mu}$ is the muon mass and $\epsilon_b$ is the binding energy of the muon in the 1s state of the muonic atom. In this case the spin projection operator coming from the conversion process is given by 

\begin{equation}
P^{(+)}_e= \frac{1}{2}(1+\mathcal{O}^e\cdot\hat{n}_0^e) 
\end{equation}
and 
\begin{align}
& \mathcal{O}^e\cdot\hat{n}_0^e= \beta \vec{\sigma}\cdot \hat{n}_0^e+ (1-\beta)(\vec{\sigma}\cdot \hat{p}_e)(\hat{p}_e\cdot \hat{n}_0^e), \\ 
& \hat{n}_0^e = (0,1,0),\quad  \hat{p}_e = (\sin \theta_s,\cos \theta_s,0).
\end{align}
Finally the wave function describing the polarized outgoing electron --coming from the conversion of a polarized muon-- is obtained by applying $P^{(+)}_e$ to the solution \eqref{electronout} and a  direct computation shows  (for a detector placed at a fixed radius R):
\begin{align}
&  \omega_{\text{conv}}(\cos \Phi >0 )- \omega_{\text{conv}}(\cos \Phi < 0 ) = R^2 \int d \Omega \cdot \text{sgn}( \hat{s}_{\mu} \cdot(\hat{p}_e\times\hat{s}_{e}) )\cdot      \psi_{as} ^{\dagger}P^{(+)}_e \psi_{as} \nonumber \\ 
 &= \frac{1}{2} G_F^2\sin \theta_s \Re e [e^{i(\delta_{-1}-\delta_{+1})}(C_R-C_L)((C_R^*+C_L^*))]  =  G_F^2\sin \theta_s \Im m (C_LC_R^*) +\mathcal{O}(\alpha 
 Z) \nonumber \\ & +\mathcal{O}(\frac{m_e}{E_e}),
\end{align}
since in the high energy limit the Coulomb phases satisfy 
\begin{equation}
 \delta_{-1}-\delta_{+1} = \frac{\pi}{2}  +\mathcal{O}(\frac{\alpha Z}{E_e}).
\end{equation}
The Coulomb phases $\delta_{\pm1}$ are defined in Eq. \eqref{deltak} and $d \Omega$ is given by $d \Omega = d\Psi d\Phi \sin \Phi$.

\acknowledgments

I would like to thank  G.~Senjanovi\'c for suggesting me the subject,  his continuous interest, advice and encouragement during the completion of this work. 
Thanks are due  to G.~Senjanovi\'c and  S. Bertolini  for enlightening and  useful  discussions and to G.~Senjanovi\'c, S. Bertolini and  A. Melfo  for a careful reading of the manuscript.  

%\paragraph{Note added.} This is also a good position for notes added
%after the paper has been written.


\end{document}